\documentstyle[preprint,aps]{revtex}
\begin{document}
\draft
\title{Momentum-Dependent Charge Transfer Excitations in $Sr_2CuO_2Cl_2$
--Angle
Resolved Electron Energy Loss Spectroscopy}
\author{Y. Y. Wang$^1$,  F. C.  Zhang$^{2,3}$, V. P. Dravid$^1$, 
 K. K.  Ng$^{2}$, M. V. Klein$^4$, S. E. Schnatterly$^5$, and L. L. Miller$^6$}
\address{
$^1$Department of Materials Science and Engineering, and Science and
Technology Center for Superconductivity, Northwestern University, Evanston,
Illinois 60208\\
$^{2}$Department of Physics, University of Cincinnati, Cincinnati, OH 45221\\
$^{3}$Department of Physics, Hong Kong University of Science and Technology\\
Clear Water Bay, Kowloon, Hong Kong\\
$^4$Department of Physics, Science and Technology Center for
Superconductivity, University of Illinois, Urbana, Illinois 61801\\
$^5$Department of Physics, University of Virginia, Charlottesville, Virginia
22901\\
$^6$Ames Laboratory, Iowa State University, Ames, Iowa 50011\\}
Content-Length: 20931
Content-Type: text
X-Lines: 423
Status: RO

\maketitle
\begin{abstract} 
Electron-hole pair excitations in the insulating cuprates $Sr_2CuO_2Cl_2$ were
investigated by angle-resolved electron energy loss spectroscopy.  The
optically allowed and optically forbidden transitions were observed to be
strongly anisotropic in $Cu-O_2$ plane. The former show a large energy
dispersion $\sim 1.5$eV along [110], and the latter appear at a higher energy
position ($\sim 4.5$eV) only along [100], but not along [110]. We interpret
these results as transitions involving excitons.  A small 
exciton model is examined to explain   both the
 observed features.
\end{abstract}


The electronic structure of the cuprate superconductors has been under intense
investigation in recent years in hope of revealing clues about the mechanism
of high $T_C$ superconductivity. While the superconductivity occurs in the
doped materials, it is of primary importance to understand the electronic
structure of the parent compounds. 

Measurements of the optical dielectric 
constant on the parent insulator reveal a
well defined peak in its imaginary part, $\epsilon_2$, at $1.6-2.0$ eV, which
measures the charge transfer band gap with excitonic correction, followed
by a weak, non-universal absorption at 1 eV higher energy\cite{Cooper}.
  Raman scattering
experiments provide additional information.  The two
magnon Raman band at $0.4$ eV energy shift shows a strong enhancement in
$YBa_2Cu_3O_{6.1}$ when the exciting laser light approaches 3.1 eV, which is
1.3 eV above the conductivity peak\cite{Cooper}.
  Similar results
were obtained from $Sr_2CuO_2Cl_2$\cite{Blumberg}.
  Optical absorption  probes the dipole allowed electron-hole
 excitations with an
essentially zero total wavevector $\vec q$.  Non-dipole allowed $q \sim 0$
excitations have been studied in the insulating cuprates through the use of
large frequency shift Raman scattering with an ultraviolet
 laser\cite{Salamon}.  Angle-resolved electron energy loss 
spectroscopy (EELS) probes
finite $\vec q$ excitations.  In the EELS experiment, the differential cross
section is directly related to the imaginary part of the inverse longitudinal
dielectric function $\epsilon (\vec q, \omega)$ by $d^2 \sigma /(dE
d\Omega) \sim q^{-2} Im[-1/\epsilon (\vec q, \omega)]$.
  At very small $q$, the
EELS measures the same excitations as the optical spectroscopy. At finite
$\vec q$,
EELS can be used to probe the dispersion of the excitation and study optically
forbidden transitions \cite{Ywang1,Nucker,Ritsko,Schnatterly,Raether}.  

In the present study, we have employed momentum-transfer resolved EELS to
investigate the dispersion of the valence band excitations in $Sr_2CuO_2Cl_2$.
  For
the first time, we report an optically forbidden transition (at $\sim 4.5$
eV )and the dispersion of the optically allowed transition, both of
which are strongly anisotropic.  The optically allowed excitation is found to
have a large dispersion range of $\sim 1.5 $ eV. This is remarkable
 in view of the narrow dispersion width $0.35$ eV for a single hole in the
same compound
observed in the 
recent photoemission experiment\cite{Wells}.
Given the latter  as a fact,  the pronounced
 large dispersion observed 
in the EELS is difficult to   
 account for  within  a usual  interband transition picture.
 Here we shall provide an exciton model for the cuprate  insulator,
  which naturally gives
 a large dispersion in the
anitferromagnetic (AF) spin background and  can also explain the 
 anisotropy in the  optically
forbidden transition in the EELS.

Crystals for this work were grown from powders using a self-flux
technique described in Ref. 10. 
  The final product of $Sr_2CuO_2Cl_2$ single crystals is a
lameller crystal  which is  hygroscopic.  Transmission
electron microscopy (TEM) specimens were freshed prepared by cleaving. The
EELS  were obtained using a cold field emission TEM (Hitachi HF-2000) equipped
with a Gatan 666 parallel electron energy loss spectrometer with the sample
cooled to $- 100^0C$.  The peak width of the unscattered beam is $\sim 0.5$
eV.  The zero loss peak was removed after fitting it to an asymmetric
Lorenzian function\cite{Ywang2}. 
  Multiple scattering
up to the fifth order was removed by using the standard $f$- sum rule
procedure\cite{Ywang3}.

Figs. 1 (a) and (b) show the EELS at different momentum, $\vec q$,
 for $Sr_2CuO_2Cl_2$ along [100] and [110] respectively. There is an optically
allowed transition in both directions, whose position is at 2.8 eV for small
$q$, corresponding to the 2 eV excitation observed in optical
spectroscopy\cite{Abbemonte,Tajima,Tokura}.
 With increasing $q$ along [100], a broad transition at
about $4.5$ eV gradually appears.  As $q$ increases, the position of the
optically allowed transition along [110] disperses systematically towards
higher energy by about $1.5$eV, while the energy position along [100] is less
dispersive.  This indicates that the dipole allowed transition becomes
anisotropic at finite $\vec q$ in the $CuO_2$ plane and that 
the bandwidth of the
excitation is large.

We have also measured the intensity of the optically allowed transition
relative to the other valence excitations in the loss function.  As $q$
increases the intensity decreases rapidly along [100], while 
 it changes little along [110]. 
 This indicates that the oscillator strength
of the dipole allowed transition is also anisotropic.

Another important feature is the optically forbidden transition located at
$4.5$ eV, at finite $\vec q$.  The transition is strongly anisotropic, 
appearing
along [100], but not along [110].  This implies a certain symmetry in the
excitations.

To gain more insight, we carried out the Kramers-Kronig analysis to extract
the excitation spectrum, or $\epsilon _(\vec q, \omega)$, from the EELS data.
We used $\epsilon (\vec q, \omega =0) \approx \epsilon _0 \equiv (q=0, \omega
=0)$ where $\epsilon_0 =4.83 $ was obtained from previous optical measurements
on this material\cite{Abbemonte,Tajima}.
  In Figs. 2 we show the resultant dispersion and
the peak intensity of the optically allowed excitation, or the peak position
and the weight of $\epsilon _2$, respectively, as functions of $\vec q$. 
 The transition peak position at 2.8 eV for small $q$ in the EELS is
shifted to 2.4 eV in $\epsilon_2$\cite{Wang}. 
 The optical forbidden transition centered
at $\sim 4.5$eV in the loss function remains at the same energy position in
$\epsilon_2$.  The dispersion of the 2.4 eV excitation in the dielectric
function along [110] is similar to that in the EELS, but along [100] the
dispersion is substantially larger than in the loss function.  However, the
excitation dispersion remains anisotropic along the different orientations.
The oscillator strength of the optical allowed excitation declines quickly
along [100], while it remains almost unchanged along [110].

It is interesting to note that the oscillator strength of the optically
forbidden exciation at $\sim 4.5$ eV compensates the strength of the optically
allowed excitation. As the scattering angle increases along [100], the
strength of the forbidden excitation grows, and the strength of the allowed one
decreases.  However, along [110], there is no forbidden excitation, and the
intensity of the optical branch remains unchanged.  This compensation
 indicates that these two excitations are closely related with each other.

The most interesting result in the EELS is the large dispersion of $\sim 1.5$
 eV for
the dipole allowed transition.  This may be surprising  
because of the much
narrow  dispersion  for the single hole observed in the
 photoemission experiment \cite{Wells}. The undoped system is an AF. A hole
moving in the AF spin background is strongly affected by the spins, so
that the coherent band width is small. 
 The same is expected for a single electron
($Cu-3d^{10}$) state.
If we consider the EELS  from the independent hole and electron band 
transition, 
we would  have a 
 much smaller dispersion than that in the EELS\cite{incoherent}. 
We also note that the top of the hole band 
is at $(\pm \pi/2, \pm \pi/2)$, so is  the bottom of the electron band
as expected. Therefore the interband transition
threshold for $\vec q = (\pi, \pi)$ and  $\vec q=0$ will be the same, which 
is inconsistent with 
the observed monotonic increase of the  dispersion along [110] 
extended to near the zone
boundary.

In a conventional  semiconductor,in addition to the elecron-hole  (e-h)
continuum,  pairs of electron and hole (excitons)
 also contribute
to the EELS. The exciton 
spectrum is  below the  e-h continuum, 
and its bandwidth is narrower
than that  of the electron or hole\cite{Mattis}. The undoped 
cuprate is a Mott insulator, and the narrow  dispersion of a
hole or electron is due to the largely incoherent motion in the spin
background. In this case, a 
 pair of electron and hole,
forming a spin singlet, may move rather freely without disturbing the spins, 
 giving a  larger
dispersion in the EELS.  We shall examine an
exciton model for the cuprate insulator to show that
the EELS dispersion can be indeed large and that the anisotropic optically
forbidden transition is due to the intensity transfer of the exciton states.  

Let us consider an undoped $CuO_2$ plane. The
 ground state is a
spin-1/2 AF of $Cu^{2+}$ with one hole at each Cu-site \cite{Vaknin}. 
 The low energy charge transition  is a hole transferred from
$Cu-3d_{x^2-y^2}$ to $O-2p\sigma$ states. Due to the strong hybridization,
 a square of  $O$ hole will bind to the central 
$Cu^{2+}$ to form a spin singlet \cite{ZR}, or a formal $Cu^{3+}$. 
 We introduce an attractive Coulomb
interaction $-V$ between the  quasi-particle ($Cu^{+}$) and the quasihole
 (formal $Cu^{3+}$). 
 For simplicity we 
consider the effective hoppings of the $Cu^+$ or $Cu^{3+}$
to be small and include the nearest neighbor (n.n.)  Coulomb attraction only. 
 Namely we  study the 
 small exciton limit as illustrated in Fig. 3.  
For a given site of $Cu^{+}$, there are four different quasiholes
 in space due to
the four-fold rotational symmetry of the $CuO_2$ plane. We can construct four
 local symmetry states with $s$, $d_{x^2-y^2}$, and two types of 
$p$ waves according to their relative phases \cite{Mrice}. We now consider 
the exciton motion, primarily due to  the n.n. $Cu-O$ 
 and the $O-O$ direct
hoppings. Because each exciton involves a pair of  neighboring $Cu$ sites ,
both are spinless, the exciton can move through the lattice without disturbing
the spins,  similar to a pair of bound n.n. holes.  When the exciton moves
 in  a finite  $\vec q$, the
different local symmetry states will mix with each other.
 For each $\vec q$, there are four eigenmodes denoted
by $S$, $D$,  $P_1$ and $P_2$ according to their local symmetries in the
limit $q \rightarrow 0$, where $P_1$ and $P_2$ are the dipole active and 
dipole
inactive modes respectively. The dispersion and the symmetry of the exciton
modes determine the EELS in our theory.

In what follows we briefly summarize the basic results obtained from 
a more detailed  mathematical formalism\cite{Zhang}.
 We  parametrize the exciton hopping
 by four integrals, which are found to be the most significant based on a 
perturbation theory  in the 
atomic
limit.  Let us  denote the exciton
 hopping
integral for $\tau(\vec R) \rightarrow \tau'(\vec R')$ to be
  $t_{\tau \tau'}(\vec R - \vec R')$, where $\vec R$ is the position of the
$Cu^+$, and $\tau$ is the  position of the formal 
$Cu^{3+}$ relative to the $Cu^+$. 
 The important hopping
integrals are,
  $t_1 \equiv  t_{\hat x,\hat y}(0)$, $t_2 \equiv t_{\hat x, \hat y}(\hat x)$,
 $-t_3 \equiv t_{\hat x, - \hat
x}(\hat x)$,
 and $t_4 \equiv t_{\hat x \hat x}(0)$.  $t_3 \sim t_{pd}^2/\epsilon_p$, with
$t_{pd}$ the $Cu-O$ hopping and  $\epsilon_p$ the
atomic energy difference between the $Cu$ and $O$ hole state.  $t_3$
is  of order of hole hopping integral in the doped $Cu$ oxides.
The dynamics of  the exciton can be described by a tight-binding Hamiltonian,
which can be solved. 
 The calculated
 dispersions are plotted in Fig. 2a for a set of parameters 
to fit 
the EELS data. The S-mode has the highest energy, while the D-mode has the 
lowest. At k=0, both S and D modes are optically forbidden because of
the d-symmetry of the Cu-hole. The $P_1$ is optically active and the charge
transfer gap is $\Delta = \epsilon_p -E_s -V + t_4 - t_3$, $E_s$ is the
hybridization energy of the $Cu-O$ singlet. 
 The optical
gap observed in the  experiment is  identified as $\Delta$,
 and a sharp  peak at $\Delta$ is expected. 
This is consistent with the  optical conductivity spectra \cite{Tokura} for
$Sr_2CuO_2Cl_2$ where  a fairly sharp profile of
 the absorption peak has been observed. 
 The exciton dispersions are anisotropic. The energy of the
$P_1$ mode is monotonic along [110] diretion, and  the dispersion width is 
given by $2t_3 $, a large energy scale.

The EELS intensity of the exciton is given by
$I \sim  q^{-2} |<\Psi_{ex}|e^{i\vec q \cdot \vec r}|GS>|^2$, where 
$|\Psi_{ex}>$  and $|GS>$ are the exciton and the ground states respectively.
 We use  the 
dipole approximation to calcaulte the intensity, so that
$I =I_0 |<\Psi_{ex}|p_1>|^2$,
where $I_0$ is the dipole active intensity of the $P_1$ mode at q=0, and
$|p_1>$ is the $P_1$ state in the limit $\vec q \rightarrow 0$.  It is
the overlap  amplitude between the exciton state and the dipole mode $p_1$
to determine the dipole active intensity in the EELS.
 The calculated  intensities of the
 excitons are plotted in
 Fig. 2b.
Along [100], the intensity of the
 $P_1$ mode is transferred to  the $S$ mode as $q$ increases, 
 which explains the
gradual appearance of the broad forbidden transition at 4.5 eV\cite{Smode}.
 Along [110] direction,
 the intensity of the 
$P_1$ mode is quite flat, and the S-mode remains dipole inactive. It can be
further shown that along [110] the S-mode does not couple
 to $p_1$, and the intensity vanishes up to the quadrupole order \cite{Zhang}.
 The 
strong anisotropy is due to the four fold lattice symmetry. These features are
in excellent agreement with the EELS data. 

The $D$ mode exciton, expected from the theory, is another optically forbidden
state. The current EELS does not reveal this mode, probably due to the limit
of the experiment resolution. Further spectroscopy works are needed to verify
this d-wave like  state.  This mode can be active in the phonon-assisted
optical process, and should be observable in the luminescence experiment.  A
recent optical absorption measurement indicates a very weak absorption state
at about 0.5 eV in the undoped cuprates \cite{Dpeak}.  The d-symmetry state
is an alternative to the magnon state or crystal field
exciton\cite{Dpeak,Sawatzky}
 proposed previously
for this weak absorption where the phonons could be involved. 

We now briefly comment on the relation to the e-h continuum, which  starts
  at
$\Delta_{e-h} = \epsilon_p - E_s + E_{kin}$, where $E_{kin}<0$ is
 the
kinetic energy of the independent 
$Cu^+$ and $Cu^{3+}$.  In our  model, $E_{kin}$ is treated 
perturbatively.    For the cuprates,  $-E_{kin}$ is $\sim 1
- 2$ eV, but there is no direct experimental data for V. The large dispersion
of the EELS reported here may be viewed as an indication that V is not very
small for the undoped materials.  If  $V \sim -E_{kin}$, which may
be the case for the cuprates, a complete exciton theory  should
also include its  spatial extension, but some  features
examined here should  remain. In this case, 
we expect  the e-h continuum starts
above the $q=0$ $P_1$ mode, and the exciton spectra  
extend into the e-h continnum due to the  large spectral dispersion. 
The  states above the e-h continuum  will then be
damped, but can exist as  resonant states contributing to the EELS.
This feature is very  different from the usual semiconductors, and is due to
the AF spin background. 
  Since the contributions to the EELS from  the  e-h continuum
have  much less  $\vec q$ dependence, we argue the excitons be dominant for
 the $\vec q$ dependence in the
EELS. 

In summary, we reported $\vec q$- resolved EELS  in
$Sr_2CuO_2Cl_2$, and observed a large energy dispersion of ~1.5 eV of the
optically allowed transition  and a very  anisotropic
optically forbidden transition at a
higher energy. We have presented a small  exciton model, in which the
exciton moves almost freely in the AF  spin  background, to
explain both features of the EELS. While more sophisticated theories are
needed for a complete description (such as the life time of the excitons and
the interaction between the excitons and the e-h continuum), 
our model contains interesting
ingredients for the seemingly unusual EELS data.    

We would like to thank N. Bulut, P. D. Han, K. Zhang, T. K. Ng, S. Haas, and 
 in particular T. M . Rice, for many useful discussions.
 The research is supported in part
 by the NSF-DMR-91-20000 through the Science and Technology Center for
Superconductivity and by the U.S. Department of Energy to Ames Laboratory
under
the contract No. W-7405-Eng-82. One of us (KKN) wishes to acknowledge the URC
at University of Cincinnati for the summer Fellowship.

\begin{figure}
\caption{Electron energy loss spectra of $Sr_2CuO_2Cl_2$ along [100] (a) and
[110] (b) with different momentum transfers, given by the inserted numbers
($\AA^{-1}$).  All intensities are in arbitrary units.}
\end{figure}

\begin{figure}
\caption{The dispersion (a) and the intensity (b) along [100] and [110]
directions. The circles are the optically allowed transitions and the squares
are the optically forbidden transitions
obtained from the EELS  after the
Kramers-Kronig analysis. The solid lines are the theoretical results for the
parameters $t_1 =0.4$, $t_2 =0.126$, $t_3=0.85$, $t_4=0.65$ eV. The energy 
for the $P$ mode at k=0 is set equal to the optically allowed
transition in the EELS, which also corresponds to the optical gap.}
\end{figure}

\begin{figure}
\caption{The structure of a local exciton in $CuO_2$ plane. The open circles
represent O-stoms, solid circles represent $Cu$ atoms, the arrows represent
spins of holes. The quasiparticle 
($Cu^+$) is at the vacant $Cu$ site $\vec R$, and the quasihole
 is on the square of $O$ atoms
 and forms a spin singlet with the central 
$Cu$ hole.}  
\end{figure}



\begin{references}
\bibitem{Cooper}S. L. Cooper $et. al.$, Phys. Rev. B{\bf 47}, 8233 (1993).
\bibitem{Blumberg}G. Blumberg,Phys. Rev. B{\bf 53}, 11930 (1996).
\bibitem{Salamon}D. Salamon $et. al.$, Phys. Rev. B{\bf 51}, 6617 (1995).
\bibitem{Ywang1}Y. Y. Wang $et. al. $,
 Phys. Rev. Lett.{\bf 75}, 2546 (1995). 
\bibitem{Nucker}N. Nucker $et. al. $, Phys. Rev. B {\bf 44}, 7155 (1991).
\bibitem{Ritsko}J. J. Ritsko $et. al. $ Phys. Rev. Lett.{\bf 36}, 210 (1976).
\bibitem{Schnatterly}S. E. Schnatterly, in Solid State Physics, Vol. {\bf
14}, Academic Press, New York (1979).
\bibitem{Raether}H. Raether, Excitation of Plasma and Interband Transitions
by Electrons, Springer-Verlag, Berlin (1980).
\bibitem{Wells}B. O. Wells $et. al. $, Phys. Rev. Lett. {\bf 74}, 964 (1995).
\bibitem{Miller}L. L. Miller $et. al. $, Phys. Rev. B {\bf 41}, 1921 (1990).
\bibitem{Ywang2} Y. Y. Wang $et. al. $, Ultramicroscopy {\bf 33}, 151 (1990).
\bibitem{Ywang3}Y. Y. Wang, Ultramicroscopy {\bf 33}, 151 (1990).
\bibitem{Abbemonte}P. Abbemonte, J. Graybeal, D. Tanner, and A. Zibold,
private communication.
\bibitem{Tajima}S. Tajima $et. al. $, Physica C {\bf 168}, 117 (1990).
\bibitem{Tokura}Y. Tokura $et. al. $, Phys. Rev. B{\bf 41}, 11657 (1990).
\bibitem{Wang}The  energy difference between the EELS and the optical
spectrum is most likely due to the instrumental broadening.  
\bibitem{incoherent}The incoherent states in both valence and
conduction bands are of large energy range, but their contributions to 
the transition are expected to be weakly $q$-dependent because of the
incoherent nature. 
\bibitem{Mattis}D. C. Mattis and J. P. Gallinar, Phys. Rev. Lett. {\bf 53},
1391 (1984).
\bibitem{Vaknin}D. Vaknin $et. al. $, Phys. Rev. B{\bf 41}, 1926 (1990).
\bibitem{ZR}F. C. Zhang and T. M. Rice, Phys. Rev. B{\bf 37}, 3759 (1988).
\bibitem{Zhang}F. C. Zhang $et. al. $, unpublished (1996).
\bibitem{Mrice}M. J. Rice and Y. R. Wang, Phys. Rev. B{\bf 36}, 8794 (1987).
\bibitem{Smode}Close examination indicates two features, one at $4.1$ eV,
 and one  at $4.9$eV. We
tentatively assign the lower one to be the $S-$ mode of the exciton discussed
in the paper. 
\bibitem{Dpeak}J. D. Perkins $et. al.$, Phys. Rev. Lett.{\bf 71}, 1621 1993.
\bibitem{Sawatzky}J. Lorentzana and G. A. Sawatzky, Phys. Rev. Lett. {\bf
74}, 1867 (1995).
\end{references}
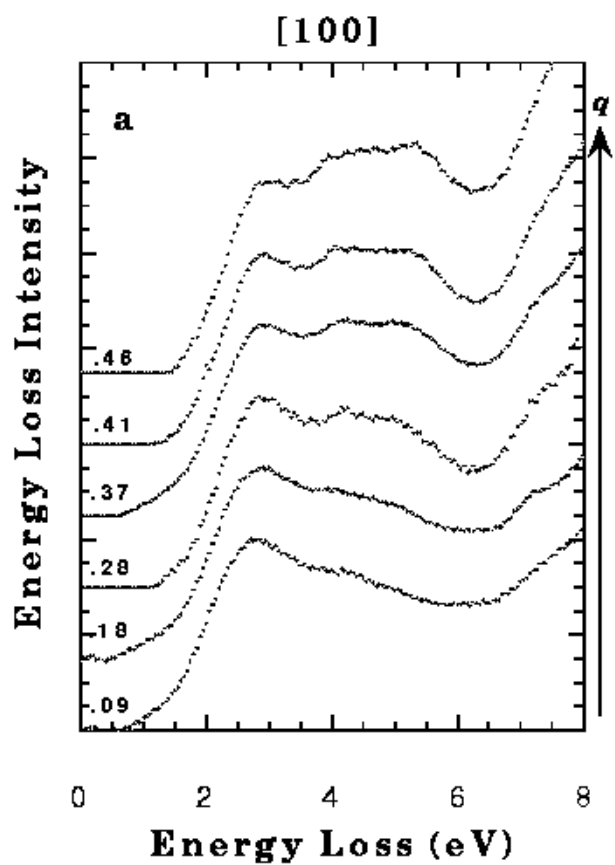
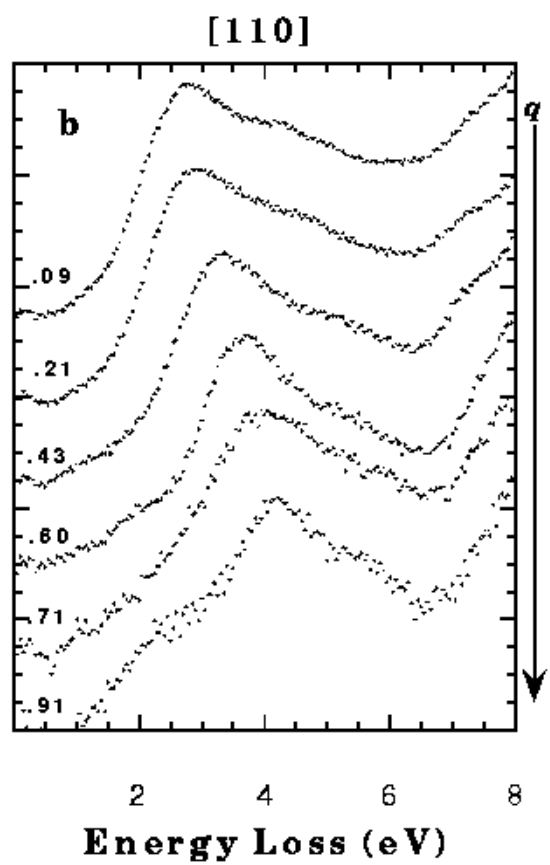
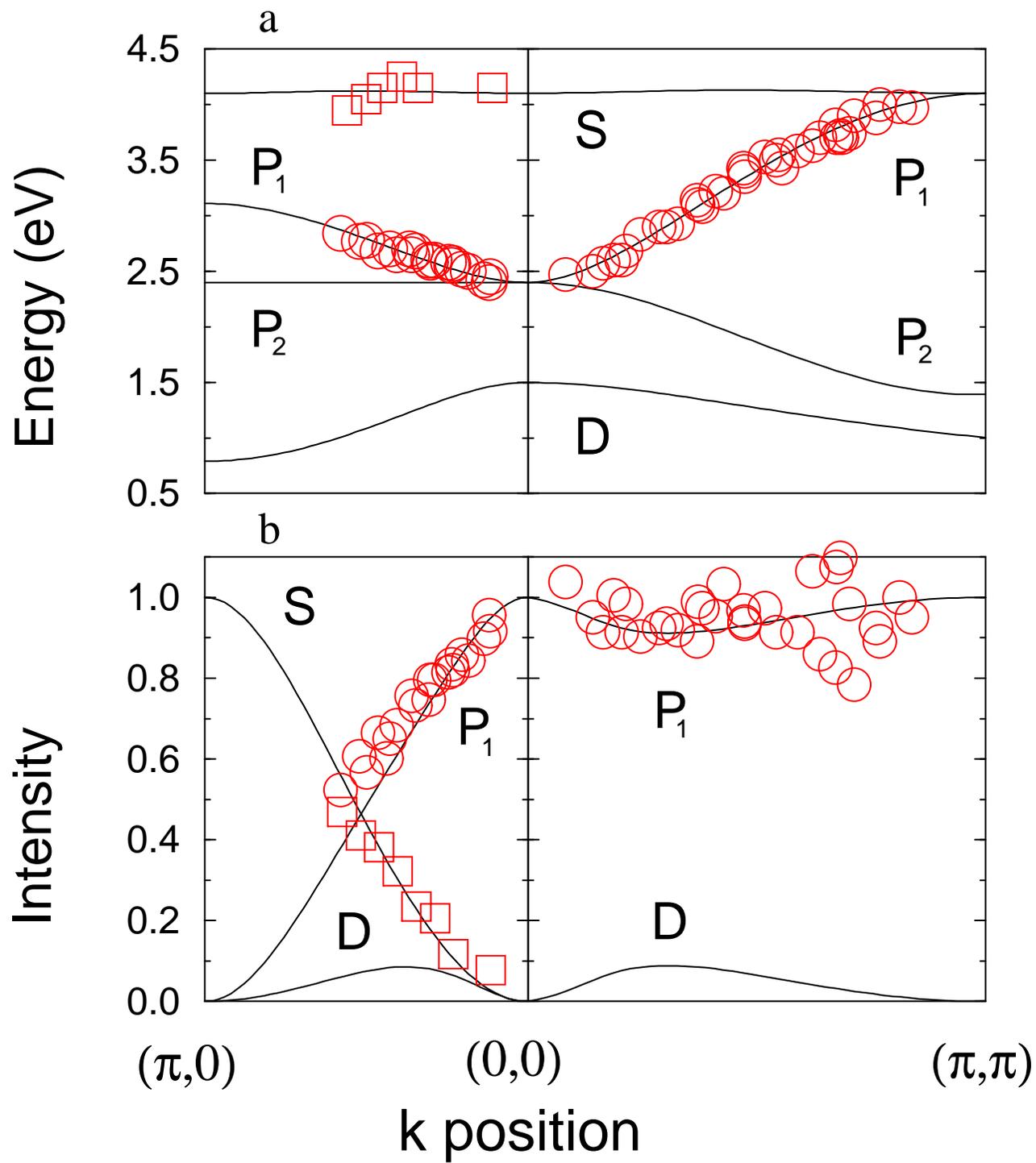
\end{document}